# Intensive Preprocessing of KDD Cup 99 for Network Intrusion Classification Using Machine Learning Techniques

[1]Ibrahim Obeidat, [2]Nabhan Hamadneh, [3]Mouhammd Alkasassbeh, [4]Mohammad Almseidin

[1, 2] Prince Al-Hussein Bin Abdullah II Faculty of Information Technology, Hashemite University, Jordan,
e-mails: { imsobeidat, nabhan}@hu.edu.jo

[3]Computer Science Department, Princess Sumaya University for Technology, Jordan.
e-mail: m.alkasassbeh@psut.edu.jo

[4]Department of Information Technology, University of Miskolc, H-3515 Miskolc, Hungary
e-mails: alsaudi@iit.uni-miskolc.hu

*Abstract*— Network security engineers work to keep services available all the time by handling intruder attacks. Intrusion Detection System (IDS) is one of the obtainable mechanism that used to sense and classify any abnormal actions. Therefore, the IDS must be always up to date with the latest intruder attacks signatures to preserve confidentiality, integrity and availability of the services. The speed of the IDS is very important issue as well learning the new attacks. This research work illustrates how the Knowledge Discovery and Data Mining (or Knowledge Discovery in Databases) KDD dataset is very handy for testing and evaluating different Machine Learning Techniques. It mainly focuses on the KDD preprocess part in order to prepare a decent and fair experimental data set. The techniques J48, Random Forest, Random Tree, MLP, Naïve Bayes and Bayes Network classifiers have been chosen for this study. It has been proven that the Random forest classifier has achieved the highest accuracy rate for detecting and classifying all KDD dataset attacks, which are of type (DOS, R2L, U2R, and PROBE).

1. INTRODUCTION

Building a reliable network is a very difficult task considering all different possible types of attacks. Nowadays, computer networks and their services are widely used in industry, business and all arenas of life. Security personnel and everyone who has a responsibility for providing protection for a network and its users, have serious concerns about intruder attacks.

Network administrators and security officers try to provide a protected environment for users' accounts, network resources, personal files and passwords. Attackers may behave in two ways to carry out their attacks on networks; one of these ways is to make a network service unavailable for users or violating personal information. Denial of service (DoS) is one of the most frequent cases representing attacks on network resources and making network services unavailable for their users. There are many types of DoS attacks, and every type has it is own behavior on consuming network resources to achieve the intruder's aim, which is to render the network unavailable for its users [1]. Remote to user (R2L) is one type of computer network attacks, which an intruder sends set of packets to another computer or server over a network where he doesn't have permission to access as a local user. User to root attacks (U2R) is a second type of attack where the intruder tries to access the network resources as a normal user and after several attempts the intruder becomes as a full access user [2]. Probing is a third type of attack in which the intruder scans network devices to determine weakness in topology design or some opened ports and then use them in the future for illegal access to personal information. There are many examples that represent probing over a network, such as nmap, portsweep, ipsweep.

Intrusion detection system (IDS) become essential part for building computer network to capture these kinds of attacks in early stages, because IDS works against all intruder attacks. IDS uses classification techniques to make decision about every packet pass through the network whether it is a normal packet or an attack (i.e. DOS, U2R, R2L, PROBE) packet.

Knowledge Discovery in Databases (KDD) is an online repository dataset, which includes all types of intruders' attacks, such as DOS, R2L, U2R, and PROBE. In this report, a number of classifiers will be evaluated on the KDD dataset. The methodology followed in this study is first to perform a preprocessing step on KDD dataset and after that using the prepared dataset on a fair environment and resources. Finally, examining which classifier is more accurate than others in detecting all studied attacks (DOS, R2L, U2R, and PROBE).

The remainder of this work is organized as follows; related work is presented in Section 2, which also provides brief discussion about KDD dataset and selected classifiers. Section 3 gives detailed steps of the preprocessing approach performed on the KDD dataset. The used classification techniques are explained in section 4. Experiments and classifiers evaluation are presented in Section 5. Section 6 presents a comprehensive comparison between the selected classifiers and experimental results with statistical values, followed by conclusions and future work in Section 7.

## 2. RELATED WORK

IDS combines hardware and software to detect attacks on networks in order to ensure the protection of the system from unauthorized access. IDS can be divided into two main classification based on the attack's detection method. The first one is the misuse and the second is anomaly detection. The anomaly detection can be used in different ways in order to detect any strange behavior of the user within the network traffic.

IDS built on Artificial Neural Network (ANN) and fuzzy clustering (FC) has been proposed to find out some networks problems and attacks. However, there is limitations of this proposed model for example, it has a lack of accuracy in low-frequent attacks. The researchers here they took over this limitation by dividing heterogeneous training set into homogeneous training subsets, by reducing the complexity of each sub-training set they reduced the complexity, the performance of detection is increased and the backup of the system can be taken successfully by using restore point [3].

Artificial intelligence technique with heuristic algorithm such as: Genetic Algorithm (GA) and ANN are used in IDS gaining its ability to learning and development, which makes them more accurate and efficient in facing the increasing number of unpredictable attacks. GA and ANN combined approach gives the IDS with extra performance and accuracy [4].

In the work of Pradhan et al [5], they took into account the user actions as a parameter in anomaly detection using a back propagation in their test. Their work very promising. The back propagation neural network had a classification rate of 100 % . the detection rate was 88% on attacks in general whether known or unknown attacks. The main advantage of this work is the minimum amount of trained data that need to give a good results of classification the traffic.

Recently, an improvement alternative of ANN is proposed called Multi-Layer Perception (MLP) ANN. The MLP method made ANN IDS methods more accurate and efficient in terms of detection and normal communication. The MLP-ANN method shows detection result much better than traditional methods. MLP overcomes the limitation of detection low frequency attacks. In addition, MLP-ANN IDS can define the type of attacks and classify them. This feature, allows system to predefine actions against similar future attacks [6] [7] [8].

In the classifier selection model presented by HuyAnh Nguyen and Deokjai Choi [9], they extracted 49,596 instances of KDD dataset and compared a set of classifiers under control environment. Kamlesh Lahreet et al [10] researchers presented different approaches to deal with KDD dataset, supervised and unsupervised methods simulated using matlab, and researchers test supervised and unsupervised techniques with fuzzy rules to identifying performance of proposed system. LEO BREIMAN [11] focused on random forest and how is it combined between trees predictors, researcher proposed error in random forest as limit number of trees in the forest.

Bhargava et al [12] in illustrated in decision tree analysis on j48 algorithm and how it is important to calculate entropy and information gain for each attributes in any dataset ready to be classified, they used decision tree with univariate and multivariate methods also researchers presented multivariate method as linear machine method. The researchers recommended this approach for large amount of data.

Chris Fleizach et al in [13] stated that nature of dataset sometimes makes it difficult to select appropriate attributes to learn, researchers implement experiments with Naïve Bayes classifier and measure performance for each call.

## 3. KNOWLEDGE DISCOVERY DATASET PREPROCESSING

MIT Lincoln labs provided KDD dataset[1], it is very helpful to examine which classifier demonstrates high accuracy to detect (DOS, R2L, U2R, and PROBE) attacks. KDD dataset has imported to Oracle database server, because there was a need to extract fairly experimental dataset for a set of classifiers with statistical information about each type of attack at KDD dataset, also to collect statistical information about each attack type instance. Table 1 illustrates KDD dataset after importing it to the database server and the table also lists number of instances for each type of attack.

TABLE 1. Number of instances for each type of attack

| Attack Type | Number of instances |
|---|---|
| SMURF(DOS) | 2,807,886 |
| NEPTUNE(DOS) | 1,072,017 |
| Back (DOS) | 2,203 |
| POD (DOS) | 264 |
| Teardrop (DOS) | 979 |
| Buffer overflow (U2R) | 30 |
| Load Module (U2R) | 9 |
| PERL (U2R) | 3 |
| Rootkit (U2R) | 10 |
| FTP Write (R2L) | 8 |
| Guess Passwd (R2L) | 53 |
| IMAP(R2L) | 12 |
| MulitHop (R2L) | 7 |
| PHF (R2L) | 4 |
| SPY (R2L) | 2 |
| Warez client (R2L) | 1,020 |
| Warez Master (R2L) | 20 |
| IPSWEEP (PROBE) | 12,481 |
| NMAP (PROBE) | 2,316 |
| PORTSWEEP(PROBE) | 10,413 |

---

[1] http://www.ll.mit.edu/ist/ideval.

| | |
|---|---:|
| SATAN (PROBE) | 15,892 |
| Normal | 972,781 |

We have 21 types of attacks, categorized into four main groups with different number of instances and occurrences. After extract full KDD dataset, all instances of experiment are full randomized; we have the following table with 148,758 instances organized as follows (Table 2):

TABLE 2. Number of Instances after organization

| Attack Type | Number of instances |
|---|---:|
| SMURF(DOS) | 85,983 |
| NEPTUNE(DOS) | 32,827 |
| Back (DOS) | 70 |
| POD (DOS) | 10 |
| Teardrop (DOS) | 30 |
| Buffer overflow (U2R) | 10 |
| Load Module (U2R) | 2 |
| PERL (U2R) | 1 |
| Rootkit (U2R) | 5 |
| FTP Write (R2L) | 2 |
| Guess Passwd (R2L) | 10 |
| IMAP(R2L) | 4 |
| MulitHop (R2L) | 2 |
| PHF (R2L) | 1 |
| SPY (R2L) | 1 |
| Warez client (R2L) | 31 |
| Warez Master (R2L) | 7 |
| IPSWEEP (PROBE) | 382 |
| NMAP (PROBE) | 70 |
| PORTSWEEP(PROBE) | 318 |
| SATAN (PROBE) | 487 |
| Normal | 28,500 |

After preparing the KDD dataset for classification experiment techniques, the idea for the next step is to work with the most common used classifier: multilayer perception, Bayesian algorithm, trees and rules using Waikato Environment for Knowledge Analysis (weka) software.

## 4. CLASSIFICATION TECHNIQUES

### A. J48 Tree.

Decision tree first introduced by [14]. It is the most common classifier used to manage database for supervised learning that gives prediction about new unlabeled data, J48 creates Univariate Decision Trees. J48 based used attribute correlation based on entropy and information gain for each attributes [12].It has been used in many fields of study, such as data mining, machine learning, information extraction, pattern recognition, and text mining. It has many advantages; it is capable of dealing with different input data types: numeric, textual and nominal. J48 decision tree is an extension of the algorithm ID3. It has an advantage over ID3 in that it can build small trees. It follows a depth-first strategy, and a divide-and-conquer approach.

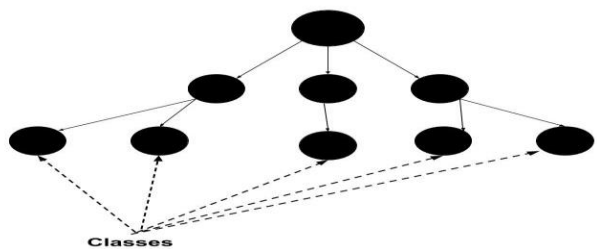

Fig. 1. Decision Tree Structure

A decision tree consists of several elements: root, internal nodes and leaves. The internal nodes represent the conditions in which the value of the parameters will be tested. Based on these values and the condition, the flow of the tree will be decided (along which branch the decision tree must go). Leaf nodes represent the decision or the class. Figure 1 shows a typical decision tree structure.

The tree is constructed by following these three main steps:

1. Ensure that all of the grouped inputs are of the same class. Then ensure that the tree is labeled with the class.
2. Calculate some parameters for each attribute, such as information gain.
3. Choose the best split attribute based on the criteria that have been set.

Entropy comes from information theory; it indicates the amount of information that is held; in other words, the higher the entropy, the more information content there is. It can be measured by:

$$\text{Entropy} = \sum_i -p_i \log_2 p_i \qquad (1)$$

Where Pi is the probability of the class 'i'.

Information gain expresses the importance of the feature or attribute, and it determines which attribute is the most important one for distinguishing between the classes to be knowledgeable. This piece of information is calculated also on training data. Information gain can help in choosing the best split; if it has a high value then this split is good, otherwise the split is not good enough. Information gain can be calculated by the data achieved from entropy:

$$\text{Information Gain} = \text{entropy (parent)} - [\text{average entropy (children)}] \qquad (2)$$

### B. Random forest

classifier was founded by LEO Breiman and Adele Cutler [14], combining tree classifiers to predict new unlabeled data. The predictor depends on a constant that denotes the number of trees in the forest; the attributes are selected randomly, and each number of set (trees) here, theu represent a one forest, and each one of these forests represents a prediction class. In this algorithm, random features selection will be selected for each individual tree.

A random forest classifier is an ensemble learning algorithm for classification and prediction of the outputs that is based on an individual number of trees [15]. Using random forest classifiers, many classification trees will be produced, and each separate tree is built by different parts of the general dataset. After each tree classifies an unlabeled class, the new object will be implemented and each tree will vote for a decision. The forest chosen as the winning class is based on the highest number of recorded votes. The number of votes is calculated as follows:

Random forest algorithms:

If there is a dataset, we need to split n samples from the whole dataset, giving (n samples= number of trees).

Each dataset sample needs to be regressed or classified; for each record this is randomly split among all predictor classes to reach an approximately optimal split. Bagging can be learned as a special scenario when m (tries) = P (number of predictors).

Predict unlabeled classes based on a reassembled number of aggregation prediction number of trees.

The accuracy rate and error rate for Random Forest are the tuning parameters for Random forest (RF) classifiers can be measured either by splitting the whole dataset, for example by testing 40% and for training 60%, or by dividing the data into 10s or 20s, etc. After a random forest is built the test model with 40% of the data can be used to calculate error rate, and accuracy rate can be measured based on comparisons of correctly classified instances with incorrectly classified instances.

Out of the bag (OOG) is another way of calculating the error rate in this technique; there is no need to split the dataset because calculation occurs within the training phase. The following parameters need to be adjusted correctly to reach the highest accuracy rate with a minimum error rate:

1- Number of trees.

2- The number of descriptors that occur randomly for present candidate's m (tries).

After analyzing and studying many cases, 500 trees are needed within the descriptor that may be desired. Even if there are great numbers of trees that will not achieve the highest accuracy rate, except for wasting training time and resources [16], random forest tuning parameters are represent a hot research area that needs to be fine-tuned. Figure 2 shows random forest architecture:

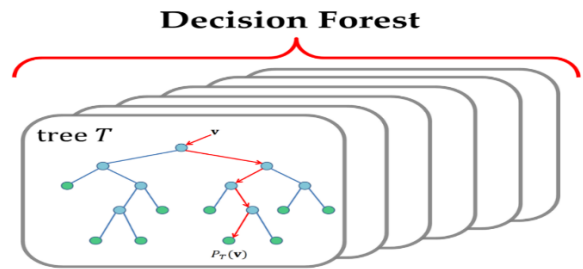

Fig. 2. Random Forest Architecture [13].

### C. Multilayer Perceptron (MLP)

MLP is widely use neural network classifier based on number of classes (output) and number of hidden layers, MLP uses weights for every node at neural network most effective attributes will get large weights conversely attributes not affect in predictive class. MLP always takes largest time for training but it has quick time for testing [17]. MLP has been positively used in daily life uses, like; regression problems, classification and prediction problems.

An example of a modest structure MLP network is illuminated in Figure 4. MLP drive the data flow to be taken in one direction from input to output. As there will be no feedback; According to [18] and [19], any MLP network can be notable by a number of performance features, which can be brief in three points:

1. Neural Network Architecture: Overall, MLP architecture can be clarified as set of links between the neurons in different layers. Generally, the architecture consists of three main layers: input layer, hidden layers and output layer. MLP is most of the time fully connected. On each link there is a weight, which is tuned based on the training algorithm.
2. Training Algorithm: is the method of selecting one model from a set of models, which tunes the weights of the links. Table 3 illustrates examples of some common transfer functions:

TABLE 3. Transfer Functions [15].

| Transfer function | definition |
|---|---|
| Linear | $f(x) = x$ |
| Sigmoid | $f(x) = \frac{1}{1+e^{-x}}$ |
| Hyperbolic | $f(x) = \frac{e^x - e^{-x}}{1+e^{-x}}$ |
| Hard limit | $f(x) = \begin{cases} 0, & x < 0 \\ 1, & x \geq 0 \end{cases}$ |
| Symmetric hard limit | $f(x) = \begin{cases} -1, & x < 0 \\ 1, & x \geq 0 \end{cases}$ |

3. Transfer Function: is applied on the net input of each neuron to control the net output signal. Here in, the function is usually non-linear. The most common function used as transfer function is Sigmoid function. The use of the sigmoid function has an advantage in neural networks trained by a back propagation learning algorithm. The sigmoid function and other common transfer functions are used as shown in Figure 3:

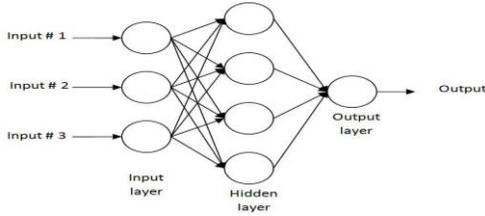

Fig. 3.  MPL Architecture

to understand how the learning process on MLP is done, here is a simple example to demonstrate the process, suppose that we have an MLP, which has N neurons as input layer and M neurons in the hidden layers, and single output neuron. The learning process will as follow:

1. Hidden layer stage: Given a number of inputs I (the output of the input layer) and a set of equivalent weights as also an input between the input and hidden neurons wij, then the outputs of all neurons in the hidden layer are calculated as in Equation (3) and Equation (4):

$$O_i = \sum_{i=0}^{N} wij \, \psi i \qquad (3)$$

$$y_j = z(O_j) \qquad (4)$$

– where i = 1, 2 . . . , N and j = 1, 2. . . M. The z and $y_j$ are the activation function and output of the $j^{th}$ node in the hidden layer, respectively. The z is usually a sigmoid function which given in Equation 5.

$$z(x) = \frac{1}{1+e^{-x}} \qquad (5)$$

2. Output stage: Equation (6) is the final outputs of all neurons in the output layer. For simplicity the equation bellow explain the output :

$$Y^\wedge = f(\sum_{j=0}^{m} wJ \, y_J^H) \qquad (6)$$

Where f() is the activation function of the output layer, which is typically a linear function. And the $Y^\wedge$ is the output of the neural network. The MLP network is always trying to make the error very small through the Back Propagation (BP) Training algorithm. At the beginning, all the weights initialized with a random values, and after that the weights are changing in each iteration until satisfied state values are obtained.

3. Error validation stage: ANN keeps learning until the error becomes very small assuming that the observed output is Y and the predicted output is ˆY. The learning process will keep going until the error difference given in Equation (7) is a minimum value, as the minimum is the best. N is the total number of instances that used during training stage.

$$Error = \frac{1}{N}\sum_{i=1}^{T}(Y_i - \hat{Y}_i)^2 \qquad (7)$$

In MLP, the weights and bias values are allocated randomly, here in, the goal of the training is to find the set of weights that give the output of the network to be close as possible to the real values.

*D. Naïve Bayes*

It is a simple probabilistic classifier that returns p (y|x), and calculates probabilistic for each class in a dataset and defines discriminative learning to predict the values of the new class. More about the main formulation for Nave Bayes may be found in [18].

A Naïve classifier links the dataset attributes x∈X that are used as inputs to the class labels Z∈ {1,2,, C}, where X is the attribute space and Z is the class space. Let X = IRD where D is a real number. The Naïve classifier may be used with discrete and continuous attributes. This model is called a multi-label problem. The learning function that directly computes class $p = (\frac{y}{x})$ is called a discriminates model. The main aim is to learn the conditional class that is used for non-linear and multi-label problems. For this reason we will use Equation (8):

$$p(y|x) = \frac{p(x,y)}{p(x)} = \frac{p(\frac{x}{y})p(y)}{\sum_{y'=1}^{c} \frac{p(\frac{x}{y'})}{p(y')}} \qquad (8)$$

The Naïve classifier achieves outputs based on an argument max function that is shown in Equation (9):

$$f(x) = y'(x) = argmax\{ \left( p\left(\frac{y}{x}\right) \right) \} \qquad (9)$$

Probabilistic classifiers have the following advantages in [18]:

1. Option to reject which is used when we are uncertain of the prediction result, so the prediction result can be ignored since human effort exists.
2. Allow learning function to be changed and a combination of probability functions can be used to reach highest performance. The main issues are if the direct learning function $p = (\frac{y}{x})$ is used and the probability function is changed; there is no need to recalculate $p = (\frac{y}{x})$.
3. Balanced classes of some of the collected datasets have unbalanced classes which means that if we have one million records of normal network traffic where there is only 1 abnormal for 1000 records we can directly train the unbalanced training dataset and easily achieve an accuracy rate of 99% by just using class always = normal. To handle such problem-balanced classes, Equation (10) and Equation (11) are used.

$$Pbal(y|x) \propto = P(x,y)Pbal(y) \qquad (10)$$

$$Ptrue(y|x) \propto Ptrue(x,y)\,Ptrue(y) \propto \frac{P\left(\frac{y}{x}\right)}{Pbal(y)} Ptrue(y) \qquad (11)$$

1- Model combinations are very useful when the collected dataset contains a mix of feature types, such as if there is a collected dataset and each feature vector represents a distinguished data type (text, images, numbers, etc.) Two or more kinds of attributes using model combinations can build two or more classifiers, such as $p\left(\frac{y}{x1}\right).p\left(\frac{y}{x2}\right)$ and so on (Murphy, 2006). To combine two different information sources, Equation (12) is used:

$$P(x1, x2|xy) = P(x1/y)P(x2/y) \qquad (12)$$

*E. Bayes Network*

It is a classifier for supervised learning that uses assumptions of independent features. It uses theory of learning that represents distribution naïve Bayesian classifier. It uses various search algorithms and different quality measure methods. Bayes Network is an enhancement for Naïve Bayes [19].

A Bayesian network is very useful, because it helps us to understand the world we are modeling. BayesNet may be the best in various areas of life, where modeling a mysterious fact and in the state of decision nets, wherever it is good to make intelligent, justifiable and quantifiable decisions that will enhance performance of classification. In brief, BayesNet is helpful for diagnosis, prediction, modeling, monitoring and classification [20].

The main idea of the Bayesian classifier consists of two phases: in the first, if an agent has an idea and knows the class, in this case it can predict the values of the other features; in the second, if the agent does not have an idea or does not know the class, in this case the Bayes rule is used to predict the class given.

We used the Bayesian Network as a classifier for the following reasons:
- Probabilistic learning, which calculates clear probabilities for assumption.
- Incremental, which is a prior knowledge and possible to be added to data viewing.
- Probabilistic prediction, which can predict more than one hypothesis, weighted by the probabilities.

The theory of the Bayesian Network is shown in Equation (13), where the symbol D indicates the training data, the probability of hypothesis h.

$$P(h|D) = \frac{P(D|h)P(h)}{P(D)} \qquad (13)$$

The symbols in Equation 4 refer to:

P (h|D): posterior probability.

P (D|h): condition probability.

P (h): prior probability of h.

P (D): marginal probability of D.

## 5. PERFORMANCE EVALUATION OF THE SELECTED CLASSIFIERS

KDD dataset presents real packets focused on wired network; it has 41 features about each packet that will help to implement different classifier types. The current experiments that are performed present fair test environment because we extracted 148,758 instances from all four groups of attack (DOS, R2L, U2R, and PROBE) as training dataset, normal packets present %19 from current experiment as original KDD dataset normal packets and the highest proportion for the DOS attack with 79% from current experiment as original KDD dataset DOS packets.

For fair control comparison between different classifiers, another 60,000 independent instances were extracted from original KDD dataset as test sample and these instances fully randomized and not included in training dataset. The experiment environment applied with Weka version 3.7.12 and Intel Xeon (R) CPU E5-2680 @ 2.70GHzX4 with available RAM 8.0GB under Ubuntu 13.10 platform. Most common classifiers are used in this experiment (J48, Random forest, Random Tree, Decision Table, Multilayer Perceptron (MLP), Naïve Bayes and Bayes Network). All models and results are saved to start comprehensive study about which classifier has the highest accuracy rate to detect attacks.

## 6. EXPERIMENT EVALUATION AND RESULTS

All selected classifiers tested with 60,000 independent instances from KDD dataset and all test instances are fully randomized. This section illustrates all parameters values that have been used in selected classifiers in the experiments.

J48 tree classifier has been tested with the parameters bellow:

Confidence factor = 0.25; numFolds = 3; seed = 1; unpruned = False, collapse tree = true, and sub tree rising =true. Random forest classifier also tested with the following parameters: Number of trees =100 and seed =1.

Random tree classifier was tested with the following parameters: Min variance = 0.001 and seed = 1. Decision table classifier was tested with the following parameters: Search techniques best first and cross value = 1. Multilayer Perceptron (MLP) classifier was tested with the following parameters: Search learning rate=0.3, momentum =0.2 and validation threshold=20. Bayes Network classifier was tested with the following parameters: Search techniques estimator value = simple estimator and search technique =K2 algorithm.

Table 4 lists statistical values that achieved in our experiments and it can be seen that random forest classifier achieves the highest Kappa statistic with rate equals to 0.8957 and the lowest Kappa statistic with Bayes network classifier with rate eqauals to 0.8464.

Table 5 records weighted average for true positive (TP) and false positive ( FP) for each classifier selected for experiment, also the random forest achieves the highest TP rate with value equals to 0.938.

Table 6 presents accuracy rate that recorded in the experiment. The random forest classifier achieves the highest rate accuracy.

TABLE 4. Statistical Values

| Classifier | Kappa statistic | Mean absolute error | Root mean squared error |
|---|---|---|---|
| J48 | 0.8844 | 0.0059 | 0.0763 |
| Random Forest | **0.8957** | 0.01 | 0.0682 |
| MLP | 0.8639 | 0.0075 | 0.0813 |
| Naïve Bayes | 0.8542 | 0.0076 | 0.0872 |
| Bayes Network | 0.8464 | 0.0085 | 0.087 |

TABLE 5. Weighted average for true positive(TP) and false positive ( FP)

| Classifier | TP Rate | FP Rate | Precision | ROC Area |
|---|---|---|---|---|
| J48 | 0.931 | 0.005 | 0.989 | 0.969 |
| **Random forest** | **0.938** | **0.001** | **0.991** | **0.996** |
| Random tree | 0.906 | 0.001 | 0.992 | 0.953 |
| MLP | 0.919 | 0.014 | 0.978 | 0.990 |
| Naïve Bayes | 0.912 | 0.002 | 0.988 | 0.969 |

TABLE 6. Accuracy rate

| Classifier | Correctly classified Instances | incorrectly classified Instances | Accuracy |
|---|---|---|---|
| J48 | 55865 | 4135 | 93.1083 % |
| **Random Forest** | **56265** | **3735** | **93.775 %** |
| Random tree | 54345 | 5655 | 90.575 % |
| MLP | 55141 | 4859 | 91.9017 % |
| Naïve Bayes | 54741 | 5259 | 91.235 % |
| Bayes Network | 54439 | 5561 | 90.7317 % |

7. CONCLUSIONS AND FUTURE WORK

Due to the urgent demand for an effective IDS in network security, researchers are striving to identify improved approaches. This research work illustrates how the KDD dataset is very useful for testing different classifiers. The work concentrates on KDD preprocess phase to prepare fair experiments and fully randomized independent test data. Among the classification techniques (J48, Random Forest, Random Tree, Decision Table, MLP, Naïve Bayes, and Bayes Network), the Random Forest classifier has achieved the highest accuracy rate for detecting and classifying all KDD dataset attack types (DOS, R2L, U2R, and PROBE). KDD dataset has 41 attributes and all of them have been recorded, but as part of future work a data engineering phase is going to be added to the study that will focus on which attribute has an essential role in achieving the highest accuracy for selected classifiers in our experiments.